# THE HST LIGHTCURVE OF (486958) 2014 $MU_{69}$

S.D. Benecchi,[1] S. Porter,[2] M.W. Buie,[2] A.M. Zangari,[2] A.J. Verbiscer,[3] K.S. Noll,[4] S.A. Stern,[2] J.R. Spencer,[2] and A. Parker[2]

[1] Planetary Science Institute, 1700 East Fort Lowell, Suite 106, Tucson, AZ 85719; susank@psi.edu
[2] Southwest Research Institute, 1050 Walnut St., Suite 300, Boulder, CO 80302.
[3] University of Virginia, Department of Astronomy, PO Box 400325, Charlottesville, VA 22904
[4] NASA Goddard Space Flight Center, 8800 Greenbelt Rd. Code 693, Greenbelt, MD 20771

Pages: 12
Figures: 8
Tables: 2

**Proposed running head:** HST Lightcurve of (486958) 2014 $MU_{69}$

**Corresponding author:** S. D. Benecchi, susank@psi.edu, Planetary Science Institute, 1700 East Fort Lowell, Suite 106, Tucson, AZ 85719.

## ABSTRACT

We report *Hubble Space Telescope* (HST) lightcurve observations of the *New Horizons* spacecraft encounter Kuiper Belt object (KBO) (486958) 2014 $MU_{69}$ acquired near opposition in July 2017. In order to plan the optimum flyby sequence the *New Horizons* mission planners needed to learn as much as possible about the target in advance of the encounter. Specifically, from lightcurve data, encounter timing could be adjusted to accommodate a highly elongated, binary, or rapidly rotating target. HST astrometric (Porter et al. 2018) and stellar occultation (Buie et al. 2018) observations constrained MU69's orbit and diameter (21 - 41 km for an albedo of 0.15 - 0.04), respectively. Photometry from the astrometric dataset suggested a variability of ≥0.3 mags, but they did not determine the period or provide shape information. To that end we strategically spaced 24 HST orbits over 9 days to investigate rotation periods from approximately 2-100 hours and to better constrain the lightcurve amplitude. Until *New Horizons* detected MU69 in its optical navigation images beginning in August 2018, this HST lightcurve campaign provided the most accurate photometry to date. The mean variation in our data is 0.15 magnitudes which suggests that MU69 is either nearly spherical (a:b axis ratio of 1:1.15), or its pole vector is pointed near the line of sight to Earth; this interpretation does not preclude a near-contact binary or bi-lobed object. However, image stacks do conclude that MU69 does not have a binary companion ≥2000 km with a sensitivity to 29$^{th}$ magnitude (an object a few km in size for an albedo of 0.04-0.15). Our data are not of sufficient signal to noise to uniquely determine the period or amplitude, however, they did provide the necessary information for spacecraft planning. We report with confidence that MU69 is not both rapidly rotating *and* highly elongated (which we define as a lightcurve amplitude ≥0.5 magnitude). Since this paper is being published post fly-by, we note that our results are consistent with the fly-by imagery and orientation of MU69 (Stern et al. 2019). The combined dataset also suggests that within the KBO lightcurve literature there are likely other objects which share a geometric configuration like MU69 resulting in an underestimate of the contact binary fraction for the Cold Classical Kuiper Belt.

# 1. INTRODUCTION

(486958) 2014 $MU_{69}$ "Ultima Thule" (hereafter MU69) is one of ~3000 objects that have thus far been identified and cataloged since discovery of the first Kuiper Belt object in 1992 (Jewitt & Luu 1993). MU69 resides in the Cold Classical region of the Kuiper Belt with a semi-major axis of $a$=44.08 AU, a nearly circular orbit with e=0.035, and low inclination of $i$=2.4°. MU69 was one of 5 objects discovered through a deep, directed search for a post-Pluto fly-by target for the *New Horizons* spacecraft (Stern et al. 2018) using the *Hubble Space Telescope* (HST) in the summer of 2014 [GO-13663; PI J. Spencer; see also Buie et al. (2018)]. Follow-up astrometric measurements (GO-14485, GO-14629, and GO-15158; PI M. Buie) confirmed its suitability for flyby accessibility, and refined its orbit for spacecraft targeting (Porter et al. 2018). Along with positional information we also acquired photometry, although at differing signal-to-noise ratios (S/N) depending on the observational geometry dictated by the need for orbit refinement. All of the data were collected using the Wide Field Camera 3 (WFC3) in the F350LP filter in order to collect as many photons as possible on the object. Exposures for the astrometric images were 368 seconds in duration.

While the *New Horizons* flyby provided a close, detailed understanding of MU69, precise encounter planning and navigation depended on the ability to characterize its physical properties from the confines of the Earth and Earth's orbit. HST has ideal capabilities to accomplish these flyby precursor support tasks. One of the key physical characteristics to investigate was MU69's shape and/or binary nature. If MU69 were significantly elongated, then the New Horizons team would want to time the close encounter to take place when the largest cross-section faces the spacecraft. Likewise, if the object were found to be rotating rapidly, then that rotation period could have influenced the instrument sequencing and integration times.

# 2. PHOTOMETRY FROM ASTROMETRIC DATASET

Our search for rotational variation began by using photometry extracted from the HST astrometric dataset (Table 1, see section 4 for details of photometry extraction). Since the data were collected near both points of quadrature as well as at opposition, the data span a wide range of S/N. We geometrically correct the photometry to an H-magnitude, $H_{F350LP}$, following the relations of Bowell et al. (1989) modified for a linear phase function:

$$H_{F350LP} = m_{F350LP}(1,1,\alpha=0^{\circ}) = mF_{350LP} - 5\log(r\Delta) - \alpha\beta. \qquad (1)$$

In Equation 1, $r$ and $\Delta$ (in AU) are the heliocentric and geocentric, distances, respectively, $\alpha$ is the solar phase angle $\beta$ is the solar phase coefficient. We consider the linear phase function appropriate because of the low the S/N of the data. Our data cover a phase range from 0.05 to 1.29 degrees and we find a phase coefficient of $\beta$ = 0.18±0.01 mag/° (Figure 1) determined by a linear fit to the non-phased, but geometrically corrected, data combining all of the measurements available. The resulting, corrected, photometry for MU69 is plotted in Figure 2.

INSERT TABLE 1 HERE
INSERT FIGURE 1 HERE
INSERT FIGURE 2 HERE

We used this astrometric dataset to place constraints on the amplitude of possible lightcurves because the wide spacing of the astrometric dataset is not suited for rotation-period determination. The uncertainties in this dataset are sizable, with a mean of 0.18 magnitudes, but sometimes as large as 0.32 magnitudes. Magnitude variations within individual HST orbits average around 0.15 magnitudes, but are as large as 0.4 magnitudes and not always systematic. Given the large range of S/N in this dataset we reasoned that variations of ≥0.3 magnitudes provided an approximate lower limit for the lightcurve amplitude of MU69 to be expected in a more densely sampled survey. This is a lower limit because assuming the rotation period is longer than 3.2 hours (2 consecutive HST orbits), a reasonable assumption for KBOs (Romanishin & Tegler 1999), then at best we are only observing part of the rise or fall in the lightcurve amplitude during the astrometric observations.

To place our assumptions in context, we note that the average rotation period for KBOs, with some degeneracy with respect to single vs. double peaked lightcurve interpretations, ranges from 7 to 9 hours (Duffard 2009; Benecchi & Sheppard 2013; Thirouin et al. 2016). However, periods as short as 3.9 hours (Haumea; Rabinowitz et al. 2006) and as long as 154 hours (Pluto-Charon; Walker & Hardie 1955) have been measured, so the 99 sparsely-measured points spaced over 4 years do not provide sufficiently dense sampling. We used the range of observed periods as a guide when designing our survey.

## 3. NEW OBSERVATIONS

Armed only with the knowledge that the total flux from MU69 varies, we set out to design an HST program to attempt to extract its rotation period and amplitude. Without prior information about how short or long the rotation period might be, we designed a program (GO-14627; PI S. Benecchi) to sample periods ranging from a few to multiple tens of hours. We assumed that the object is more likely elongated rather than spherical given our knowledge of small KBO lightcurves thus far (Trilling et al. 2006). If we measure large amplitude (>0.4 magnitudes) variation then we will be able to more reliably estimate MU69's true rotation. If, however, the lightcurve amplitude is small, then there is likely to be some ambiguity in the true period determination (Harris et al. 2014). In either case our experiment was designed to provide useful constraints for spacecraft planning.

Our measurement sequence utilized 4 visits of 6 HST orbits for a total of 24 orbits within the timespan of 224.66 hours (based on exposure mid-times; 9.36 days), near opposition, between June 25 and July 4, when MU69 should be brightest. Visit 1 spanned 10.12 hours and was separated from Visit 2 by 13.7 hours (0.57 days). Visits 2 spanned 11.65 hours, and was separated from visit 3 by 34.3 hours (1.4 days). Visit 3 spanned 11.71 hours and was separated from visit 4 by 131.3 hours (5.47 days). Visit 4 spanned 11.7 hours. The orbits within a visit were as consecutive as possible, in order to minimize aliasing interpretations of the period; however, HST gyroscope limitations require the telescope to move off the MU69 field after 3-4 consecutive orbits. Therefore, all of the visit orbit sequences have a one or two orbit gap someplace within them. Using synthetic lightcurve datasets we concluded prior to scheduling that these gaps would not significantly impact the ability to interpret our results. The HST fields were examined in advance for background sources to make sure that we optimized the photometric return and minimized contaminated observations. Data were collected in the F350LP filter to obtain the highest S/N possible (as good as S/N~7 in the actual dataset) with integrations of 367 seconds in duration, allowing for 5 exposures in each HST orbit.

## 4. PHOTOMETRY FROM LIGHTCURVE DATASET

Photometry was carried out with an IDL PSF-Tiny Tim (Krist & Hook 2004) matching routine which uses an amoeba (Press et al. 1992) downhill simplex method minimization to match the model to the data by iteratively adjusting the values of the fitted parameters until the $\chi 2$ converged (Benecchi et al. 2009). Due to the low S/N of the data ~10 iterations, which we checked as the conversion happened to make sure that the model was not stuck on any background variations, were typically required to reach a final PSF model.

Initially each image was modelled with both a single and double PSF (fitting for x1,y1, flux1 in one case and x1,y1,flux1,x2,y2,flux2 in the second case) in addition to fitting for the sky background. Likewise, we also fit single and double PSFs to the orbit, visit, and campaign stacked images to look for a binary companion.

Next we ran a photometric analysis where instead of deriving the (x,y) position from the images themselves, we forced the (x,y) position on each image to be that projected by the best orbit solution for MU69 (Porter et al. 2018). In this case we fit for only the object flux and sky background. This seemed the best way to consistently extract the fluxes for all of the HST data available for MU69 independent of the original purpose for the observations, especially since some of the measurements in the astrometric campaign were barely detectable.

The calibrated fluxes (Table 2), expressed in the ST-magnitude (STMAG) and Vega-magnitude (VEGAMAG) systems, were derived from the observed counts matched to the actual MU69 images for an infinite aperture on the Tiny Tim models using the inverse sensitivity and photometric zero point keyword values (PHOTFLAM and PHOTZPT) from the HST image headers (Rajan et al. 2011). Photometry is extracted from the model data to remove noise from background sources that increase the background signal and respective uncertainty in the raw images. To estimate uncertainty on the flux itself we re-run the data/model comparison with steps in flux to determine the value at which the $\chi^2$ residual of the image changes by 1-sigma. Figure 3 shows the results in F350LP Vega-magnitudes for each of the individual lightcurve campaign visits. For reference this Vega-magnitude system (the reference system for most ground-based measurements) is 0.321 magnitudes brighter than the ST-magnitude system (http://www.stsci.edu/hst/wfc3/analysis/uvis_zpts/uvis1_infinite).

INSERT TABLE 2 HERE
INSERT FIGURE 3 HERE

## 5. DATA ANALYSIS

### *5.1. Binary Evaluation*

MU69 was easily identified in individual exposures; in most cases there were no obvious background sources or nearby bright stars. From the results of the binary vs. single image PSF fits, we conclude that MU69 is not a resolved binary in the individual HST images.

Additionally we stacked the images within each orbit to search for fainter companions that might have been missed in the individual image analysis. Images from each HST orbit (5 images per orbit, with orbit 12 having only 4 images due to an HST timing constraint), each visit (30 images over 6 orbits), and from the entire lightcurve campaign (119 images) were stacked to search for faint companions (Figure 4); no faint companions were identified. Therefore, we can say with confidence that MU69 does not have a binary companion separated by ≥2000km, with a sensitivity to 29$^{th}$ magnitude. This limit corresponds to an object a few km in size for an assumed albedo of 0.15-0.04.

INSERT FIGURE 4 HERE

### *5.2. Lightcurve Extraction*

We ran two sets of analyses on the extracted magnitudes from our survey data: one considered data from only the new lightcurve campaign (119 images in total spanning 224.66 hours (9.36 days) and a second considered all of the photometry available for MU69 (an additional 99 images made since 2014), yielding a 4-year baseline (Figure 1, bottom). To put all these data on the same baseline, we geometrically corrected all of the data (Table 2) following the same process as that described in Section 2. The caveat for this second analysis is that the uncertainties at some epochs are large, so when we fit a period to the data we weight the points by their uncertainties and when we evaluate the amplitude of a potential lightcurve we only consider the lightcurve campaign data since they have both high signal to noise as well as the appropriate time sampling for lightcurve work. All of our analysis is done using light-time corrected times-stamps derived from the observation Julian date mid-times from the images corrected for the observing geometry as provided in Tables 1 & 2.

We analyze the data using a modified Phase Dispersion Minimization (PDM; Stellingwerf 1978) fitting technique that goes through every possible period and folds the data, then fits a second-order Fourier series to each folded lightcurve (Buie et al. 2018); hereafter referred to as the "Fourier PDM" technique. This modeling is different from the traditional PDM which bins the data and looks for the place where the points in the bins are not as dispersed as other periods, but it has the advantage of being able to deal with data that is very sparsely sampled. For our dataset we searched a range of periods between 2 and 100 hours (single-peaked; 4-200 hours double-peaked) then focused on better-sampled periods between 2-30 (4-60) hours. We used a step interval of 1000000 which is dependent upon the search range, but at the lowest resolution samples periods every 30 seconds. We are suspicious of periods near to 3.2 hours because these are commensurate with double the HST sampling rate (96 minutes), however we do not exclude them from our testing range.

As a check on our methodology, we also ran a Lomb-Scargle analysis (Scargle 1982; Press & Rybicki 1989) on our datasets to look for consistency between different period estimation techniques. This technique should work for the 9-day 2017 lightcurve campaign, but is not designed for the extremely sparse sampling of the astrometric dataset.

In the Fourier PDM the "goodness of fit" parameter is a chi-square measurement so the lower this value the better the fit. In the Lomb-Scargle model the "goodness of fit" parameter is measured as a peak in the periodogram so the higher the value the more likely the result. For the lightcurve campaign (looking from 2-10, 2-40 and 2-100 hours; Figure 5) using both search algorithms we see what appear to be best-fit values at a single-peaked period of 3.4 hours, double-peaked period of 6.8 hours, however compared to the range of the plot, the marginal significance compared to other possible values is low. The Lomb-Scargle actually gives a slight preference for a period of 7.4 (14.8 double-peaked) hours. For the Fourier PDM the chi-square range is 0.35 units and the difference between the minimum and the average is only 1-sigma from the scatter. In the Lomb-Scargle model the range is 4.8, but there are clearly other peaks with not dis-similar significances near 10.8, 40 and even 90 hours. If we analyze each of the 6-orbit individual lightcurve campaign segments with either of these two models we get similarly inconclusive results.

For the astrometric dataset alone if we allow short periods we mostly find peaks slightly less than 4 hours (8 hours) which should not be commensurate with the HST orbit period since these data are mostly single HST orbit datasets. If we exclude periods shorter than 4 hours (based on an assumed break-up rotational speed barrier; Romanishin & Tegler 1999) we still find the best fit period to be near the lower limit of the search range. If we combine the datasets we find a period of 3.38 hours, double-peaked period of 6.76 hours, although if we exclude periods less than 4 hours we find a period near 21.6 hours (43.2 hours; Figure 6). Since there are more data points in the lightcurve campaign than in the entire astrometric campaign, this lightcurve dataset still dominates the conclusion. We note that the astrometric campaign data does not appear to significantly contradict the results of the lightcurve campaign, however the chi-square range range is still small compared to the best resultant chi-square. Therefore, we still believe there is too much uncertainty in the dataset to decisively settle on a best-fit period.

Finally, we use the data we have to place limits on the shape of MU69 and/or the geometry of the object by looking at the range of the data during each of the 6-orbit lightcurve segments, since this is the highest signal to noise dataset. The proposed variation, ≥0.3 magnitudes, that we used as a lower limit from the astrometric dataset alone in our original analysis is primarily due to the lower S/N of the photometry for measurements outside of opposition. When we fit solely to the lightcurve dataset we find that for any period the amplitude is between 0.15 and 0.5 magnitudes. Fits to the full 4-visit lightcurve dataset yield an amplitude of 0.15-0.2 magnitudes. Unfortunately, this is still comparable to the uncertainty in the dataset. While again this is not a definitive result, it does provide a useful physical limit and allows us to determine that MU69 is not significantly elongated (like the recently discovered interstellar visitor, Bannister et al. 2017; Meech et al. 2017) or suggests that the object orientation is approximately pointed in the observing direction; in either case there is no strong argument for adjusting the timing of the spacecraft encounter for observing a particularly long axis.

INSERT FIGURE 5 HERE
INSERT FIGURE 6 HERE

### *5.3. Elongation/shape*

If MU69 is a triaxial ellipsoid with semi-major axes a≥b≥c in rotation about the c-axis, then the minimum and maximum flux of the rotation curve measured in magnitudes, Δm, can be used to determine the projection of the body shape (i.e. how spherical the object is) into the plane of the sky:

$$\Delta m = 2.5\log\frac{a}{b} - 1.25\log\left(\frac{a^2\cos^2\theta + c^2\sin^2\theta}{b^2\cos^2\theta + c^2\sin^2\theta}\right). \quad (2)$$

The parameter θ is the angle at which the rotation axis is inclined to the line of sight (an object with θ=90° is being viewed equatorially; Binzel et al. 1989). If we make the extreme assumption that we are in fact viewing the object equatorially, then this equation can be rearranged to give the axis ratio, $a/b = 10^{0.4\Delta m}$, and in that case MU69 has an axis ratio of 1.15-1.2.

### *5.4. Synthetic Modeling*

#### *5.4.1. HST Dataset Limits*

Synthetic modeling provides another approach to understanding the limits of the HST dataset. We generate synthetic lightcurves for periods ranging from short (4 hours) to long (100 hours) and with peak-to-peak amplitudes ranging from 0.2 to 0.6 magnitudes. We embed these lightcurves in the magnitude randomized HST dataset and then see how many of the synthetic lightcurves we can recover, and at what precision, using the same Fourier PDM analysis as described earlier. Figure 7 gives the results for determining the amplitude of the real object in which we find unsurprisingly that the smaller the amplitude the more difficult it is to recover it accurately. This exercise demonstrates, since we do not measure an amplitude larger than ~0.15 magnitudes in the HST lightcurve campaign data, again, the strong suggestion that MU69 is NOT significantly elongated, or that its pole geometry is along the line of sight to Earth, or we are within the phase space for ambiguous interpretations/object configurations suggested by Harris et al. (2014).

INSERT FIGURE 7 HERE

#### *5.4.2. Projected New Horizons Optical Navigation Results*

Using this same synthetic modeling, the timing and S/N of the *New Horizons* planned optical navigation (Op-Nav) and full body science images, the spacecraft data will allow us to determine the rotation period of MU69 during its 2018 approach and 2019 encounter. For a period of 5-10 hours *New Horizons* should be able to resolve the period for an amplitude of 0.05 magnitudes or larger to within 0.01 hours with 85% certainty. For a longer period of ~20 hours *New Horizons* can resolve the same amplitude to within 0.1 hours with the similar certainty. For a period of ~40 hours our accuracy drops to 50% and for a period of ~100 hours it drops to 10%, however the period can still be determined to within a one hour uncertainty if it is low amplitude. If the amplitude is at least 0.15 magnitudes, the limit of the HST data, then *New Horizons* should be able to determine any period to an accuracy of better than 0.1 hours in all cases and for periods less than ~20 hours to an accuracy of better than 0.01 hours. However, the in-situ images will allow for a clear rotational determination independent of the Op-Nav images since we can use surface features to unambiguously track MU69's rotational motion.

## SUMMARY AND MISSION COMPARISON

Despite the substantial amount of relatively high quality data acquired on MU69 using HST, we find that we cannot uniquely determine its rotation period and amplitude. The HST dataset presented here confirm with confidence that unless the pole vector is pointed close to the line of sight to Earth, MU69 is not rapidly rotating AND highly elongated (which we define as a amplitude lightcurve ≥0.5 magnitude). The data preclude the existence of a binary companion ≥2000km with a sensitivity to 29$^{th}$ magnitude (an object a few km in size for assumed albedos of 0.04-0.15). Attempts to measure and model the lightcurve amplitude yield an average of 0.15 magnitudes, comparable to the scatter in the measurements themselves. An object with such a low amplitude lightcurve would be consistent with being relatively spherical, with an axis ratio less than 1.15. One possible explanation for a small lightcurve amplitude is that HST is viewing MU69 nearly pole-on or with the line of sight nearly perfectly pointed towards Earth. If this configuration is correct, then MU69 could still be significantly elongated and present a low amplitude variation. The lightcurve period and shape remains undetermined and does not rule out the possibility of a near-contact binary, or bi-lobed, interpretation proposed by stellar occultation data (Buie et al. 2018).

Finally, since this paper will come out after the MU69 encounter with *New Horizons* we reflect on our results in the context of the fly-by images (Stern et al. 2019, in press). Op-Nav data up to the point of encounter still produced a non-unique rotational period. The first resolved image of MU69 showed it to be a bi-lobed object consistent with the occultation results. The higher resolution images acquired during the fly-by showed that it is in fact rotating in a face-on/line of sight orientation such that the lobes rotate around a common center point, but the same overall surface-area is more or less always pointed towards Earth. In other words, the spin axis of MU69 lies within the cone within which the brightness variations due to a changing cross-section is smaller than the photometric errors in our measurements (Lacerda & Luu 2003). This is one of the ambiguous results presented in Harris et al (2014) and is consistent with our lightcurve non-detection.

The fact that the object is bi-lobed also significantly impacts the lightcurve amplitude that we observed in that a fully elliptical object rotating in the same way would present a lightcurve with a comparatively larger amplitude. Using the preliminary reported pole (Porter et al, 2019,) of J2000 RA = 300° and Dec -21°, we calculate that the Earth-pole angle viewing angle of MU69 was inclined to the viewer 14° (Zangari et al, 2019). Combining this information with preliminary size measurements for the two lobes (9.73+/-0.02 km and 7.12+/-0.06 km, Stern et al 2019, Bierson et al, 2019), we can estimate the approximate light curve amplitude we might have observed. Under the assumption that both objects are completely spherical, there is no “neck,” and that the light reflected from each object is exactly proportional to the visible object area, Figure 8 shows the lightcurve for several pole inclinations with the preliminary 14° estimate highlighted giving an idealized lightcurve amplitude of 0.003 magnitudes. For comparison, a projected ellipsoid model (Connolley & Ostro 1984) with identical edge-on maxima and minima to the contact binary would have an amplitude of 0.019 magnitude, a factor of 6 larger. Unfortunately, both of these values are well below the minimum detectable lightcurve amplitude of 0.15 magnitudes reported herein and even below any feasible photometry accuracy for an object of this magnitude from any ground-based or Earth-orbit facility. While these models are both simplistic, for a lightcurve amplitude as large as the scatter in our data, 0.15 magnitudes, to be produced would require a contact binary pole inclination of ≥52° or an ellipsoidal pole inclination of ≥40°. This observation suggests that within the KBO lightcurve literature there are probably other objects which share a geometric configuration like MU69 resulting in an underestimate of the contact binary fraction for the Cold Classical Kuiper Belt.

INSERT FIGURE 8 HERE

## ACKNOWLEDGMENTS

Observations were made with the NASA/ESA Hubble Space Telescope, obtained at the Space Telescope Science Institute, which is operated by the Association of Universities for Research in Astronomy, Inc., under NASA contract NAS 5-26555. These observations are associated with programs GO-14627, GO-13663, GO-14485, GO-14629, and GO-15158. Support was provided by NASA through a grant from the Space Telescope Science Institute, which is operated by the Association of Universities for Research in Astronomy, Inc., under NASA contract NAS 5-26555. The modeling efforts in this work were supported by the National Aeronautics and Space Administration under Grant/Contract/Agreement No. NNX15AE04G issued through the SSO Planetary Astronomy Program. Some coauthors acknowledge support from the NASA New Horizons Project.

## REFERENCES

Bannister, M. T., and 12 colleagues. 2017. Col-OSSOS: Colors of the Interstellar Planetesimal 1I/'Oumuamua. The Astrophysical Journal 851, L38.

Benecchi, S. D., & Sheppard, S. S. 2013. Light Curves of 32 Large Transneptunian Objects. The Astronomical Journal, 145(5), 124. doi:10.1088/0004-6256/145/5/124

Benecchi, S. D., Noll, K. S., Grundy, W. M., Buie, M. W., Stephens, D. C., & Levison, H. F. 2009. The Correlated Colors of Transneptunian Binaries. Icarus, 200, 292. doi:10.1016/j.icarus.2008.10.025

Bierson et al. 2019. LPSC 50.

Binzel, R. P., Farinella, P., Zappala, V., & Cellino, A. 1989. In*: Asteroids II; Proceedings of the Conference*, ed. R. P. Binzel, T. Gehrels, and M. S. Matthews (Univ. of Arizona Press, Tucson), 416-441.

Bowell, E., Hapke, B., Domingue, D., et al. 1989, in Asteroids II, ed. R. P. Binzel, T. Gehrels, & M. S. Matthews (Tucson, AZ: Univ. Arizona Press), 524-556.

Buie, M. W, Zangari, A. M., Marchi, S., Levison, H. F. and Mottola, S. 2018. Light Curves of Lucy Targets: Leucus and Polymele 155, 245. doi:10.3847/1538-3881/aabd81

Buie, M., and 11 colleagues 2018. Pre-encounter update on (486958) 2014MU69 and occultation results from 2017 and 2018. AAS/Division for Planetary Sciences Meeting Abstracts 509.06.

Connelly, R. & Ostro, S. J. 1984. Ellipsoids and lightcurves, NASA STI/Recon Technical Report A, 17, 87.

Duffard, R., Ortiz, J.L., Thirouin, A., Santos-Sanz, P., Morales, N., 2009. Transneptunian objects and Centaurs from light curves. arXiv. doi:10.1051/0004-6361/200912601

Harris, A. W., Pravec, P., Galád, A., Skiff, B. A., Warner, B. D., Világi, J., Gajdoš, Š., Carbognani, A., Hornoch, K., Kušnirák, P., Cooney, W. R., Gross, J., Terrell, D., Higgins, D., Bowell, E. & Koehn, B. W. 2014. On the maximum amplitude of harmonics of an asteroid lightcurve, Icarus, 235, 55.

Jewitt, D., & Luu, J. 1993. Discovery of the candidate Kuiper belt object 1992 QB1. Nature (ISSN 0028-0836), 362, 730–732.

Krist, J., Hook, R., 2004. The Tiny Tim User's Guide: Version 6.3. STScI, Baltimore; available at: http://www.stsci.edu/software/tinytim/

Lacerda, P. & Luu, J. 2003. On the detectability of lightcurves of Kuiper belt objects, Icarus, 161, 174.

Meech, K. J., and 17 colleagues 2017. A brief visit from a red and extremely elongated interstellar asteroid. Nature 552, 378.

Moore, J. M., McKinnon, W. B., Cruikshank, D. P., Gladstone, G. R., Spencer, J. R., Stern, S. A., et al. 2018. Great expectations: Plans and predictions for new horizons encounter with Kuiper Belt object 2014 MU69 (“Ultima Thule”). Geophysical Research Letters, 45. https://doi.org/10.1029/2018GL078996

Porter et al. 2019. A Contact Binary in the Kuiper Belt: The Shape and Pole of (486958) 2014 $MU_{69}$. LPSC 50

Porter, S. B., Buie, M. W., Parker, A. H., Spencer, J. R., Benecchi, S., Tanga, P., Verbiscer, A., Kavelaars, J. J., Gwyn, S. D. J., Young, E. F., Weaver, H. A., Olkin, C. B., Parker, J. W., and Stern, S. A.. 2018. High-precision Orbit Fitting and Uncertainty Analysis of (486958) 2014 MU69. The Astronomical Journal 156, 20.

Press, W.H., Teukolsky, S.A., Vetterling, W.T., Flannery, B.P., 1992. Numerical Recipes in C, second ed. Cambridge University Press, New York.

Press, W.H. and Rybicki, G.B. 1989. Fast algorithm for spectral analysis of unevenly sampled data. ApJ 338, 277–280. doi:10.1086/167197.

Rabinowitz, D. L., Barkume, K., Brown, M. E., Roe, H., Schwartz, M., Tourtellotte, S. & Trujillo, C. 2006. Photometric Observations Constraining the Size, Shape, and Albedo of 2003 EL61, a Rapidly Rotating, Pluto-sized Object in the Kuiper Belt, The Astrophysical Journal, 639, 1238.

Rajan, A. et al. 2011 "WFC3 Data Handbook", Version 2.1, (Baltimore: STScI). http://www.stsci.edu/hst/wfc3/documents/handbooks/currentDHB/wfc3_dhb.pdf

Romanishin, W., Tegler, S.C., 1999. Rotation rates of Kuiper-belt objects from their light curves. Nature 398, 129. doi:10.1038/18168

Scargle, J.D., 1982. Studies in astronomical time series analysis. II - Statistical aspects of spectral analysis of unevenly spaced data. Astrophys. J. 263, 835–853. doi:10.1086/160554

Stellingwerf, R. F. 1978. The Astrophysical Journal, 224, pp. 953-960.

Stern, S. A. et al. 2019. Overview of Initial Results From the Reconnaissance Flyby of a Kuiper Belt Planetesimal: 2014 $MU_{69}$. LPSC 50.

Stern, S.A., Weaver, H. A., Spencer, J. R., Elliot, H. A. and the New Horizons Team (2018). The New Horizons Kuiper Belt Extended Mission. Space Sci Rev (2018) 214:77; https://doi.org/10.1007/s11214-018-0507-4

Thirouin, A., Sheppard, S. S., Noll, K. S., Moskovitz, N. A., Ortiz, J. L., Doressoundiram, A. 2016. Rotational Properties of the Haumea Family Members and Candidates: Short-term Variability. The Astronomical Journal 151, 148.

Thirouin, A., Ortiz, J. L., Duffard, R., et al. 2010, *A&A*, 522, 93.

Thirouin, A., Sheppard, S. 2018. Lightcurves of the dynamically cold classical transneptunian objects. DPS 50, 302.05. Knoxville, TN.

Trilling, D. E. & Bernstein, G. M. 2006. Light Curves of 20-100 km Kuiper Belt Objects Using the Hubble Space Telescope, The Astronomical Journal, 131, 1149.

Walker, M. F. & Hardie, R. 1955. A Photometric determination of the rotational period of Pluto, Publications of the Astronomical Society of the Pacific, 67, 224.

Zangari et al. 2019. The Mysterious Missing Light Curve of (486958) 2014 $MU_{69}$, A Bi-lobate Contact Binary Visited by New Horizons. LPSC 50.

# TABLES

**TABLE 1. PHOTOMETRY FROM HST ASTROMETRY CAMPAIGNS**

| Image Rootname | JD (midtime) | Light-Time Corrected JD (midtime) | ST_mag | Vega_mag | Magerr | R (au) | Δ (au) | α(°) | 1-wayLT | $H_{ST_mag}$ |
|---|---|---|---|---|---|---|---|---|---|---|
| icii11r7q_1 | 2456834.86924 | 2456834.62440 | 27.17 | 26.85 | 0.14 | 43.41 | 42.40 | 0.148 | 352.570 | 10.43 |
| icii11raq_1 | 2456834.88153 | 2456834.63669 | 27.43 | 27.11 | 0.17 | 43.41 | 42.40 | 0.148 | 352.570 | 10.69 |
| icii11rcq_1 | 2456834.88768 | 2456834.64284 | 27.33 | 27.01 | 0.15 | 43.41 | 42.40 | 0.148 | 352.570 | 10.59 |
| icii11req_1 | 2456834.89382 | 2456834.64898 | 27.41 | 27.09 | 0.15 | 43.41 | 42.40 | 0.148 | 352.570 | 10.66 |
| icii12rqq_1 | 2456835.00812 | 2456834.76328 | 27.13 | 26.81 | 0.13 | 43.41 | 42.40 | 0.145 | 352.570 | 10.39 |
| icii12rsq_1 | 2456835.01426 | 2456834.76942 | 27.46 | 27.14 | 0.17 | 43.41 | 42.40 | 0.145 | 352.570 | 10.72 |
| icii12ruq_1 | 2456835.02041 | 2456834.77557 | 27.38 | 27.06 | 0.16 | 43.41 | 42.40 | 0.145 | 352.570 | 10.64 |
| iciig7cwq_2 | 2456872.04538 | 2456871.79980 | 27.54 | 27.22 | 0.18 | 43.41 | 42.53 | 0.676 | 353.627 | 10.69 |
| iciig7cyq_2 | 2456872.05152 | 2456871.80595 | 27.41 | 27.09 | 0.17 | 43.41 | 42.53 | 0.676 | 353.627 | 10.57 |
| iciig7d0q_2 | 2456872.05767 | 2456871.81209 | 27.16 | 26.84 | 0.13 | 43.41 | 42.53 | 0.676 | 353.627 | 10.32 |
| iciig7d2q_2 | 2456872.06381 | 2456871.81824 | 27.43 | 27.11 | 0.17 | 43.41 | 42.53 | 0.676 | 353.627 | 10.58 |
| iciig8kaq_2 | 2456873.23983 | 2456872.99420 | 27.03 | 26.71 | 0.12 | 43.41 | 42.54 | 0.699 | 353.710 | 10.18 |
| iciig8kcq_2 | 2456873.24598 | 2456873.00035 | 27.41 | 27.09 | 0.16 | 43.41 | 42.54 | 0.699 | 353.710 | 10.56 |
| iciig8keq_2 | 2456873.25212 | 2456873.00649 | 27.81 | 27.49 | 0.22 | 43.41 | 42.54 | 0.699 | 353.710 | 10.96 |
| iciig9ryq_2 | 2456890.83129 | 2456890.58456 | 27.86 | 27.54 | 0.24 | 43.41 | 42.73 | 1.002 | 355.298 | 10.95 |
| iciig9s0q_2 | 2456890.83744 | 2456890.59070 | 27.50 | 27.18 | 0.18 | 43.41 | 42.73 | 1.002 | 355.298 | 10.59 |
| iciig9s2q_2 | 2456890.84358 | 2456890.59685 | 27.43 | 27.11 | 0.16 | 43.41 | 42.73 | 1.002 | 355.298 | 10.52 |
| iciih0s4q_2 | 2456890.88535 | 2456890.63861 | 27.65 | 27.33 | 0.21 | 43.41 | 42.73 | 1.002 | 355.306 | 10.74 |
| iciih0s5q_2 | 2456890.89150 | 2456890.64476 | 27.64 | 27.32 | 0.20 | 43.41 | 42.73 | 1.003 | 355.306 | 10.73 |
| iciih0s7q_2 | 2456890.89764 | 2456890.65090 | 27.53 | 27.21 | 0.18 | 43.41 | 42.73 | 1.003 | 355.306 | 10.61 |
| iciih0s9q_2 | 2456890.90379 | 2456890.65705 | 27.45 | 27.13 | 0.16 | 43.41 | 42.73 | 1.003 | 355.306 | 10.54 |
| iciih0sbq_2 | 2456890.90994 | 2456890.66320 | 27.23 | 26.91 | 0.15 | 43.41 | 42.73 | 1.003 | 355.306 | 10.32 |
| iciih3byq_2 | 2456892.67695 | 2456892.43008 | 27.47 | 27.15 | 0.19 | 43.41 | 42.75 | 1.029 | 355.498 | 10.56 |
| iciih3bzq_2 | 2456892.68310 | 2456892.43622 | 27.38 | 27.06 | 0.18 | 43.41 | 42.75 | 1.029 | 355.498 | 10.46 |
| iciih3c1q_2 | 2456892.68924 | 2456892.44237 | 27.55 | 27.23 | 0.19 | 43.41 | 42.75 | 1.029 | 355.498 | 10.63 |
| iciih3c3q_2 | 2456892.69539 | 2456892.44851 | 27.83 | 27.51 | 0.25 | 43.41 | 42.75 | 1.029 | 355.498 | 10.91 |
| iciih3c5q_2 | 2456892.70153 | 2456892.45466 | 27.46 | 27.14 | 0.18 | 43.41 | 42.75 | 1.029 | 355.498 | 10.55 |
| iciih4c7q_2 | 2456892.74332 | 2456892.49644 | 27.27 | 26.95 | 0.16 | 43.41 | 42.75 | 1.030 | 355.506 | 10.35 |
| iciih4c8q_2 | 2456892.74946 | 2456892.50258 | 27.28 | 26.96 | 0.15 | 43.41 | 42.75 | 1.030 | 355.506 | 10.37 |
| iciih4caq_2 | 2456892.75561 | 2456892.50873 | 27.29 | 26.97 | 0.15 | 43.41 | 42.75 | 1.030 | 355.506 | 10.38 |
| iciih4ccq_2 | 2456892.76175 | 2456892.51487 | 27.81 | 27.49 | 0.23 | 43.41 | 42.75 | 1.030 | 355.506 | 10.89 |
| iciij5yfq_2 | 2456945.57313 | 2456945.32139 | 27.87 | 27.55 | 0.28 | 43.40 | 43.59 | 1.291 | 362.516 | 10.86 |
| iciij5yjq_2 | 2456945.58542 | 2456945.33368 | 27.36 | 27.04 | 0.18 | 43.40 | 43.59 | 1.291 | 362.516 | 10.35 |
| iciij6yyq_2 | 2456945.64563 | 2456945.39388 | 27.65 | 27.33 | 0.23 | 43.40 | 43.60 | 1.291 | 362.524 | 10.64 |
| iciij6z2q_2 | 2456945.65792 | 2456945.40617 | 27.33 | 27.01 | 0.18 | 43.40 | 43.60 | 1.291 | 362.533 | 10.32 |
| iciij7etq_2 | 2456946.69482 | 2456946.44297 | 27.60 | 27.28 | 0.24 | 43.40 | 43.61 | 1.285 | 362.674 | 10.59 |
| iciij7eyq_2 | 2456946.71326 | 2456946.46140 | 27.53 | 27.21 | 0.21 | 43.40 | 43.61 | 1.285 | 362.674 | 10.53 |
| iciij9c7q_2 | 2456952.87053 | 2456952.61809 | 27.49 | 27.17 | 0.22 | 43.40 | 43.71 | 1.245 | 363.514 | 10.48 |
| iciij9c9q_2 | 2456952.87667 | 2456952.62423 | 27.58 | 27.26 | 0.21 | 43.40 | 43.71 | 1.245 | 363.514 | 10.57 |
| ict101egq_2 | 2457147.19178 | 2457146.94431 | 27.51 | 27.19 | 0.18 | 43.38 | 42.85 | 1.137 | 356.362 | 10.57 |
| ict101eiq_2 | 2457147.19791 | 2457146.95043 | 27.65 | 27.33 | 0.19 | 43.38 | 42.85 | 1.137 | 356.362 | 10.71 |
| ict103vwq_2 | 2457208.11057 | 2457207.86594 | 27.30 | 26.98 | 0.17 | 43.38 | 42.36 | 0.052 | 352.271 | 10.58 |
| ict103vyq_2 | 2457208.11670 | 2457207.87207 | 27.47 | 27.15 | 0.18 | 43.38 | 42.36 | 0.052 | 352.271 | 10.75 |
| ict103w0q_2 | 2457208.12284 | 2457207.87821 | 27.30 | 26.98 | 0.14 | 43.38 | 42.36 | 0.052 | 352.271 | 10.57 |
| ict103w2q_2 | 2457208.12897 | 2457207.88434 | 27.35 | 27.03 | 0.15 | 43.38 | 42.36 | 0.052 | 352.271 | 10.63 |
| id3m01hhq_2 | 2457462.50829 | 2457462.25608 | 27.62 | 27.30 | 0.24 | 43.36 | 43.67 | 1.236 | 363.181 | 10.62 |
| id3m01hiq_2 | 2457462.51440 | 2457462.26219 | 28.01 | 27.69 | 0.32 | 43.36 | 43.67 | 1.236 | 363.181 | 11.01 |
| id3m01hsq_2 | 2457462.53273 | 2457462.28052 | 27.63 | 27.31 | 0.27 | 43.36 | 43.67 | 1.236 | 363.181 | 10.63 |
| id3m02snq_2 | 2457523.59107 | 2457523.34455 | 27.28 | 26.96 | 0.14 | 43.35 | 42.69 | 1.009 | 354.982 | 10.37 |
| id3m02soq_2 | 2457523.59718 | 2457523.35066 | 27.86 | 27.54 | 0.22 | 43.35 | 42.69 | 1.009 | 354.982 | 10.95 |
| id3m02sqq_2 | 2457523.60329 | 2457523.35678 | 27.63 | 27.31 | 0.19 | 43.35 | 42.69 | 1.009 | 354.982 | 10.72 |
| id3m02suq_2 | 2457523.61551 | 2457523.36900 | 26.76 | 26.44 | 0.11 | 43.35 | 42.69 | 1.008 | 354.982 | 9.85 |
| id5901erq_2 | 2457595.20717 | 2457594.96235 | 27.35 | 27.03 | 0.16 | 43.35 | 42.39 | 0.475 | 352.537 | 10.56 |
| id5901etq_2 | 2457595.21328 | 2457594.96846 | 27.31 | 26.99 | 0.15 | 43.35 | 42.39 | 0.475 | 352.537 | 10.51 |
| id5901evq_2 | 2457595.21939 | 2457594.97458 | 27.17 | 26.85 | 0.14 | 43.35 | 42.39 | 0.475 | 352.537 | 10.38 |
| id5901exq_2 | 2457595.22550 | 2457594.98069 | 27.86 | 27.54 | 0.24 | 43.35 | 42.39 | 0.475 | 352.537 | 11.06 |
| id5902a9q_2 | 2457682.71348 | 2457682.46169 | 27.42 | 27.10 | 0.20 | 43.34 | 43.60 | 1.271 | 362.583 | 10.42 |
| id5902adq_2 | 2457682.72570 | 2457682.47390 | 27.56 | 27.24 | 0.23 | 43.34 | 43.60 | 1.271 | 362.591 | 10.55 |
| id5902afq_2 | 2457682.73181 | 2457682.48001 | 27.78 | 27.46 | 0.26 | 43.34 | 43.60 | 1.271 | 362.591 | 10.78 |
| id5953gxq_2 | 2457875.37689 | 2457875.12932 | 27.21 | 26.89 | 0.14 | 43.32 | 42.87 | 1.191 | 356.504 | 10.26 |
| id5953gyq_2 | 2457875.38300 | 2457875.13543 | 27.29 | 26.97 | 0.13 | 43.32 | 42.87 | 1.191 | 356.504 | 10.34 |
| id5953h0q_2 | 2457875.38911 | 2457875.14154 | 27.20 | 26.88 | 0.14 | 43.32 | 42.87 | 1.191 | 356.504 | 10.25 |
| id5953h2q_2 | 2457875.39523 | 2457875.14765 | 27.30 | 26.98 | 0.14 | 43.32 | 42.87 | 1.190 | 356.504 | 10.35 |
| id5953h4q_2 | 2457875.40134 | 2457875.15376 | 27.60 | 27.28 | 0.19 | 43.32 | 42.87 | 1.190 | 356.504 | 10.65 |
| id5904wiq_2 | 2457899.22773 | 2457898.98201 | 27.48 | 27.16 | 0.15 | 43.32 | 42.55 | 0.866 | 353.826 | 10.60 |
| id5904wkq_2 | 2457899.23384 | 2457898.98812 | 27.32 | 27.00 | 0.14 | 43.32 | 42.55 | 0.866 | 353.826 | 10.44 |
| id5904wmq_2 | 2457899.23995 | 2457898.99424 | 27.41 | 27.09 | 0.15 | 43.32 | 42.55 | 0.866 | 353.826 | 10.54 |
| id5906kaq_2 | 2457985.11989 | 2457984.87404 | 27.54 | 27.22 | 0.17 | 43.31 | 42.57 | 0.920 | 354.026 | 10.65 |
| id5906kbq_2 | 2457985.12600 | 2457984.88015 | 27.42 | 27.10 | 0.16 | 43.31 | 42.57 | 0.920 | 354.026 | 10.54 |
| id5906kdq_2 | 2457985.13211 | 2457984.88626 | 27.58 | 27.26 | 0.18 | 43.31 | 42.57 | 0.920 | 354.026 | 10.69 |
| id5906khq_2 | 2457985.14434 | 2457984.89849 | 27.53 | 27.21 | 0.17 | 43.31 | 42.57 | 0.920 | 354.026 | 10.64 |
| idoy07yuq_2 | 2458054.06655 | 2458053.81447 | 27.19 | 26.87 | 0.16 | 43.31 | 43.65 | 1.235 | 362.998 | 10.20 |
| idoy07ywq_2 | 2458054.07266 | 2458053.82058 | 27.70 | 27.38 | 0.25 | 43.31 | 43.65 | 1.235 | 362.998 | 10.71 |
| idoy07yyq_2 | 2458054.07877 | 2458053.82669 | 27.03 | 26.71 | 0.15 | 43.31 | 43.65 | 1.235 | 362.998 | 10.04 |
| idoy07z0q_2 | 2458054.08488 | 2458053.83280 | 27.67 | 27.35 | 0.25 | 43.31 | 43.65 | 1.235 | 362.998 | 10.67 |
| idoy08idq_2 | 2458193.59001 | 2458193.33794 | 27.91 | 27.59 | 0.28 | 43.30 | 43.65 | 1.223 | 362.982 | 10.92 |
| idoy09jdq_2 | 2458259.27628 | 2458259.03028 | 27.42 | 27.10 | 0.15 | 43.29 | 42.60 | 0.975 | 354.242 | 10.52 |
| idoy09jfq_2 | 2458259.28239 | 2458259.03639 | 27.39 | 27.07 | 0.15 | 43.29 | 42.60 | 0.975 | 354.242 | 10.49 |
| idoy10ktq_2 | 2458321.12211 | 2458320.87783 | 27.40 | 27.08 | 0.16 | 43.29 | 42.30 | 0.315 | 351.772 | 10.64 |
| idoy10kvq_2 | 2458321.12822 | 2458320.88394 | 26.99 | 26.67 | 0.11 | 43.29 | 42.30 | 0.315 | 351.772 | 10.23 |
| idoy10kxq_2 | 2458321.13433 | 2458320.89005 | 27.61 | 27.29 | 0.19 | 43.29 | 42.30 | 0.315 | 351.772 | 10.85 |
| idoy10kzq_2 | 2458321.14045 | 2458320.89616 | 27.05 | 26.73 | 0.12 | 43.29 | 42.30 | 0.315 | 351.772 | 10.29 |
| idoy11f5q_2 | 2458356.89321 | 2458356.64716 | 27.60 | 27.28 | 0.17 | 43.28 | 42.61 | 0.999 | 354.317 | 10.71 |
| idoy11f7q_2 | 2458356.89932 | 2458356.65327 | 27.82 | 27.50 | 0.20 | 43.28 | 42.61 | 1.000 | 354.317 | 10.92 |
| idw612qsq_2 | 2458430.30320 | 2458430.05042 | 26.61 | 26.29 | 0.13 | 43.28 | 43.77 | 1.136 | 364.005 | 9.63 |
| idw612qvq_2 | 2458430.31543 | 2458430.06265 | 27.35 | 27.03 | 0.21 | 43.28 | 43.77 | 1.136 | 364.005 | 10.36 |

**TABLE 2. PHOTOMETRY FROM HST LIGHTCURVE CAMPAIGN**

| Image Rootname | JD (midtime) | Light-Time Corrected JD (midtime) | ST_mag | Vega_mag | Magerr | R (au) | Δ (au) | α(°) | 1-wayLT | $H_{ST_mag}$ |
|---|---|---|---|---|---|---|---|---|---|---|
| id8i01skq | 2457929.81763 | 2457929.57322 | 27.01 | 26.69 | 0.15 | 43.32 | 42.32 | 0.250 | 351.947 | 10.65 |
| id8i01slq | 2457929.82375 | 2457929.57934 | 26.82 | 26.50 | 0.15 | 43.32 | 42.32 | 0.250 | 351.947 | 10.46 |
| id8i01snq | 2457929.82986 | 2457929.58545 | 27.13 | 26.81 | 0.15 | 43.32 | 42.32 | 0.250 | 351.947 | 10.77 |
| id8i01spq | 2457929.83597 | 2457929.59156 | 26.92 | 26.60 | 0.15 | 43.32 | 42.32 | 0.250 | 351.947 | 10.56 |
| id8i01srq | 2457929.84208 | 2457929.59767 | 26.68 | 26.36 | 0.15 | 43.32 | 42.32 | 0.250 | 351.947 | 10.32 |
| id8i02stq | 2457929.88385 | 2457929.63944 | 26.93 | 26.61 | 0.15 | 43.32 | 42.32 | 0.249 | 351.947 | 10.57 |
| id8i02suq | 2457929.88996 | 2457929.64555 | 27.00 | 26.68 | 0.15 | 43.32 | 42.32 | 0.249 | 351.947 | 10.64 |
| id8i02swq | 2457929.89607 | 2457929.65166 | 26.63 | 26.31 | 0.15 | 43.32 | 42.32 | 0.249 | 351.947 | 10.27 |
| id8i02syq | 2457929.90218 | 2457929.65777 | 26.83 | 26.51 | 0.15 | 43.32 | 42.32 | 0.249 | 351.947 | 10.47 |
| id8i02t0q | 2457929.90829 | 2457929.66388 | 26.73 | 26.41 | 0.15 | 43.32 | 42.32 | 0.248 | 351.947 | 10.37 |
| id8i03t4q | 2457929.95006 | 2457929.70566 | 27.17 | 26.85 | 0.16 | 43.32 | 42.32 | 0.247 | 351.938 | 10.81 |
| id8i03t7q | 2457929.95618 | 2457929.71178 | 26.84 | 26.52 | 0.16 | 43.32 | 42.32 | 0.247 | 351.938 | 10.48 |
| id8i03t9q | 2457929.96229 | 2457929.71789 | 26.82 | 26.50 | 0.16 | 43.32 | 42.32 | 0.247 | 351.938 | 10.46 |
| id8i03tbq | 2457929.96840 | 2457929.72400 | 26.55 | 26.23 | 0.16 | 43.32 | 42.32 | 0.247 | 351.938 | 10.19 |
| id8i03tdq | 2457929.97451 | 2457929.73011 | 27.16 | 26.84 | 0.16 | 43.32 | 42.32 | 0.247 | 351.938 | 10.80 |
| id8i04tfq | 2457930.01629 | 2457929.77189 | 26.81 | 26.49 | 0.15 | 43.32 | 42.32 | 0.246 | 351.938 | 10.45 |
| id8i04tgq | 2457930.02240 | 2457929.77800 | 27.24 | 26.92 | 0.15 | 43.32 | 42.32 | 0.246 | 351.938 | 10.88 |
| id8i04tiq | 2457930.02851 | 2457929.78411 | 26.73 | 26.41 | 0.15 | 43.32 | 42.32 | 0.246 | 351.938 | 10.37 |
| id8i04tkq | 2457930.03463 | 2457929.79023 | 26.96 | 26.64 | 0.15 | 43.32 | 42.32 | 0.245 | 351.938 | 10.60 |
| id8i04tmq | 2457930.04074 | 2457929.79634 | 26.64 | 26.32 | 0.15 | 43.32 | 42.32 | 0.245 | 351.938 | 10.27 |
| id8i05txq | 2457930.14873 | 2457929.90433 | 26.93 | 26.61 | 0.09 | 43.32 | 42.32 | 0.243 | 351.938 | 10.57 |
| id8i05tyq | 2457930.15484 | 2457929.91044 | 26.92 | 26.60 | 0.09 | 43.32 | 42.32 | 0.243 | 351.938 | 10.56 |
| id8i05u0q | 2457930.16096 | 2457929.91656 | 26.80 | 26.48 | 0.09 | 43.32 | 42.32 | 0.243 | 351.938 | 10.44 |
| id8i05u2q | 2457930.16707 | 2457929.92267 | 26.91 | 26.59 | 0.09 | 43.32 | 42.32 | 0.243 | 351.938 | 10.55 |
| id8i05u4q | 2457930.17318 | 2457929.92878 | 27.05 | 26.73 | 0.09 | 43.32 | 42.32 | 0.243 | 351.938 | 10.69 |
| id8i06u6q | 2457930.21494 | 2457929.97054 | 27.11 | 26.79 | 0.17 | 43.32 | 42.32 | 0.242 | 351.938 | 10.75 |
| id8i06u7q | 2457930.22105 | 2457929.97665 | 26.93 | 26.61 | 0.17 | 43.32 | 42.32 | 0.241 | 351.938 | 10.57 |
| id8i06u9q | 2457930.22716 | 2457929.98276 | 26.71 | 26.39 | 0.17 | 43.32 | 42.32 | 0.241 | 351.938 | 10.35 |
| id8i06ubq | 2457930.23327 | 2457929.98887 | 27.28 | 26.96 | 0.17 | 43.32 | 42.32 | 0.241 | 351.938 | 10.92 |
| id8i06udq | 2457930.23938 | 2457929.99498 | 27.02 | 26.70 | 0.17 | 43.32 | 42.32 | 0.241 | 351.938 | 10.66 |
| id8i07crq | 2457930.81090 | 2457930.56651 | 27.01 | 26.69 | 0.17 | 43.32 | 42.32 | 0.229 | 351.922 | 10.65 |
| id8i07ctq | 2457930.81701 | 2457930.57262 | 26.96 | 26.64 | 0.17 | 43.32 | 42.32 | 0.229 | 351.922 | 10.60 |
| id8i07cwq | 2457930.82312 | 2457930.57873 | 26.60 | 26.28 | 0.17 | 43.32 | 42.32 | 0.228 | 351.922 | 10.25 |
| id8i07cyq | 2457930.82923 | 2457930.58484 | 27.65 | 27.33 | 0.17 | 43.32 | 42.32 | 0.228 | 351.922 | 11.29 |
| id8i07d1q | 2457930.83534 | 2457930.59095 | 26.73 | 26.41 | 0.17 | 43.32 | 42.32 | 0.228 | 351.922 | 10.38 |
| id8i08daq | 2457930.87713 | 2457930.63274 | 27.00 | 26.68 | 0.13 | 43.32 | 42.32 | 0.227 | 351.922 | 10.64 |
| id8i08dbq | 2457930.88324 | 2457930.63885 | 26.76 | 26.44 | 0.13 | 43.32 | 42.32 | 0.227 | 351.922 | 10.40 |
| id8i08ddq | 2457930.88935 | 2457930.64496 | 26.78 | 26.46 | 0.13 | 43.32 | 42.32 | 0.227 | 351.922 | 10.43 |
| id8i08dfq | 2457930.89546 | 2457930.65107 | 26.78 | 26.46 | 0.13 | 43.32 | 42.32 | 0.227 | 351.922 | 10.42 |
| id8i08diq | 2457930.90157 | 2457930.65718 | 26.66 | 26.34 | 0.13 | 43.32 | 42.32 | 0.227 | 351.922 | 10.30 |
| id8i09dkq | 2457930.94335 | 2457930.69897 | 26.71 | 26.39 | 0.14 | 43.32 | 42.32 | 0.226 | 351.913 | 10.36 |
| id8i09dlq | 2457930.94946 | 2457930.70508 | 27.00 | 26.68 | 0.14 | 43.32 | 42.32 | 0.225 | 351.913 | 10.65 |
| id8i09dnq | 2457930.95557 | 2457930.71119 | 26.64 | 26.32 | 0.14 | 43.32 | 42.32 | 0.225 | 351.913 | 10.29 |
| id8i09dpq | 2457930.96169 | 2457930.71731 | 26.98 | 26.66 | 0.14 | 43.32 | 42.32 | 0.225 | 351.913 | 10.63 |
| id8i09drq | 2457930.96780 | 2457930.72342 | 26.64 | 26.32 | 0.14 | 43.32 | 42.32 | 0.225 | 351.913 | 10.29 |
| id8i10dtq | 2457931.00957 | 2457930.76519 | 26.99 | 26.67 | 0.14 | 43.32 | 42.32 | 0.224 | 351.913 | 10.64 |
| id8i10duq | 2457931.01568 | 2457930.77130 | 26.70 | 26.38 | 0.14 | 43.32 | 42.32 | 0.224 | 351.913 | 10.35 |
| id8i10dwq | 2457931.02179 | 2457930.77741 | 26.55 | 26.23 | 0.14 | 43.32 | 42.32 | 0.224 | 351.913 | 10.19 |
| id8i10dyq | 2457931.02790 | 2457930.78352 | 26.75 | 26.43 | 0.14 | 43.32 | 42.32 | 0.224 | 351.913 | 10.40 |
| id8i10e0q | 2457931.03401 | 2457930.78963 | 26.94 | 26.62 | 0.14 | 43.32 | 42.32 | 0.224 | 351.913 | 10.58 |
| id8i11ejq | 2457931.21489 | 2457930.97051 | 26.89 | 26.57 | 0.15 | 43.32 | 42.32 | 0.220 | 351.913 | 10.53 |
| id8i11elq | 2457931.22100 | 2457930.97662 | 26.91 | 26.59 | 0.15 | 43.32 | 42.32 | 0.220 | 351.913 | 10.55 |
| id8i11eoq | 2457931.22711 | 2457930.98273 | 27.13 | 26.81 | 0.15 | 43.32 | 42.32 | 0.219 | 351.913 | 10.77 |
| id8i11eqq | 2457931.23322 | 2457930.98884 | 26.65 | 26.33 | 0.15 | 43.32 | 42.32 | 0.219 | 351.913 | 10.30 |
| id8i12esq | 2457931.27804 | 2457931.03366 | 26.95 | 26.63 | 0.14 | 43.32 | 42.32 | 0.218 | 351.905 | 10.59 |
| id8i12etq | 2457931.28415 | 2457931.03977 | 26.54 | 26.22 | 0.14 | 43.32 | 42.32 | 0.218 | 351.905 | 10.18 |
| id8i12evq | 2457931.29026 | 2457931.04588 | 26.66 | 26.34 | 0.14 | 43.32 | 42.32 | 0.218 | 351.905 | 10.31 |
| id8i12exq | 2457931.29637 | 2457931.05199 | 26.84 | 26.52 | 0.14 | 43.32 | 42.32 | 0.218 | 351.905 | 10.48 |
| id8i13lvq | 2457932.73121 | 2457932.48685 | 26.65 | 26.33 | 0.07 | 43.32 | 42.32 | 0.187 | 351.880 | 10.30 |
| id8i13lwq | 2457932.73732 | 2457932.49296 | 26.66 | 26.34 | 0.07 | 43.32 | 42.32 | 0.186 | 351.880 | 10.31 |
| id8i13lyq | 2457932.74343 | 2457932.49907 | 26.77 | 26.45 | 0.07 | 43.32 | 42.32 | 0.186 | 351.880 | 10.42 |
| id8i13m0q | 2457932.74954 | 2457932.50518 | 26.81 | 26.49 | 0.07 | 43.32 | 42.32 | 0.186 | 351.880 | 10.46 |
| id8i13m3q | 2457932.75566 | 2457932.51130 | 26.71 | 26.39 | 0.07 | 43.32 | 42.32 | 0.186 | 351.880 | 10.36 |
| id8i14m5q | 2457932.79743 | 2457932.55307 | 26.54 | 26.22 | 0.14 | 43.32 | 42.31 | 0.185 | 351.872 | 10.20 |
| id8i14m6q | 2457932.80354 | 2457932.55918 | 26.72 | 26.40 | 0.14 | 43.32 | 42.31 | 0.185 | 351.872 | 10.37 |
| id8i14m8q | 2457932.80965 | 2457932.56529 | 26.88 | 26.56 | 0.14 | 43.32 | 42.31 | 0.185 | 351.872 | 10.54 |
| id8i14maq | 2457932.81576 | 2457932.57140 | 26.88 | 26.56 | 0.14 | 43.32 | 42.31 | 0.184 | 351.872 | 10.53 |
| id8i14mcq | 2457932.82187 | 2457932.57751 | 26.75 | 26.43 | 0.14 | 43.32 | 42.31 | 0.184 | 351.872 | 10.40 |
| id8i15meq | 2457932.86363 | 2457932.61927 | 26.85 | 26.53 | 0.13 | 43.32 | 42.31 | 0.184 | 351.872 | 10.50 |
| id8i15mfq | 2457932.86974 | 2457932.62538 | 26.75 | 26.43 | 0.13 | 43.32 | 42.31 | 0.184 | 351.872 | 10.40 |
| id8i15mhq | 2457932.87585 | 2457932.63149 | 26.40 | 26.08 | 0.13 | 43.32 | 42.31 | 0.183 | 351.872 | 10.05 |
| id8i15mjq | 2457932.88196 | 2457932.63760 | 26.77 | 26.45 | 0.13 | 43.32 | 42.31 | 0.183 | 351.872 | 10.42 |
| id8i15mlq | 2457932.88807 | 2457932.64371 | 26.65 | 26.33 | 0.13 | 43.32 | 42.31 | 0.183 | 351.872 | 10.30 |
| id8i16mnq | 2457932.92986 | 2457932.68550 | 27.06 | 26.74 | 0.15 | 43.32 | 42.31 | 0.182 | 351.872 | 10.71 |
| id8i16moq | 2457932.93597 | 2457932.69161 | 26.63 | 26.31 | 0.15 | 43.32 | 42.31 | 0.182 | 351.872 | 10.28 |
| id8i16mqq | 2457932.94208 | 2457932.69772 | 26.80 | 26.48 | 0.15 | 43.32 | 42.31 | 0.182 | 351.872 | 10.45 |
| id8i16msq | 2457932.94819 | 2457932.70383 | 26.81 | 26.49 | 0.15 | 43.32 | 42.31 | 0.182 | 351.872 | 10.46 |
| id8i16muq | 2457932.95430 | 2457932.70994 | 26.85 | 26.53 | 0.15 | 43.32 | 42.31 | 0.182 | 351.872 | 10.50 |
| id8i17obq | 2457933.12850 | 2457932.88414 | 26.77 | 26.45 | 0.13 | 43.32 | 42.31 | 0.178 | 351.872 | 10.42 |
| id8i17ocq | 2457933.13461 | 2457932.89025 | 26.80 | 26.48 | 0.13 | 43.32 | 42.31 | 0.178 | 351.872 | 10.45 |
| id8i17oeq | 2457933.14072 | 2457932.89636 | 26.75 | 26.43 | 0.13 | 43.32 | 42.31 | 0.178 | 351.872 | 10.40 |
| id8i17ogq | 2457933.14684 | 2457932.90248 | 26.78 | 26.46 | 0.13 | 43.32 | 42.31 | 0.177 | 351.872 | 10.44 |
| id8i17oiq | 2457933.15295 | 2457932.90859 | 26.44 | 26.12 | 0.13 | 43.32 | 42.31 | 0.177 | 351.872 | 10.09 |
| id8i18okq | 2457933.19472 | 2457932.95036 | 26.88 | 26.56 | 0.14 | 43.32 | 42.31 | 0.176 | 351.872 | 10.54 |
| id8i18olq | 2457933.20083 | 2457932.95648 | 26.87 | 26.55 | 0.14 | 43.32 | 42.31 | 0.176 | 351.864 | 10.52 |
| id8i18onq | 2457933.20694 | 2457932.96259 | 26.92 | 26.60 | 0.14 | 43.32 | 42.31 | 0.176 | 351.864 | 10.57 |
| id8i18opq | 2457933.21305 | 2457932.96870 | 26.40 | 26.08 | 0.14 | 43.32 | 42.31 | 0.176 | 351.864 | 10.05 |
| id8i18orq | 2457933.21916 | 2457932.97481 | 26.93 | 26.61 | 0.14 | 43.32 | 42.31 | 0.176 | 351.864 | 10.59 |
| id8i19geq | 2457938.69074 | 2457938.44644 | 26.97 | 26.65 | 0.15 | 43.32 | 42.31 | 0.066 | 351.797 | 10.64 |
| id8i19gfq | 2457938.69685 | 2457938.45255 | 27.03 | 26.71 | 0.15 | 43.32 | 42.31 | 0.066 | 351.797 | 10.70 |
| id8i19gpq | 2457938.70296 | 2457938.45866 | 26.75 | 26.43 | 0.15 | 43.32 | 42.31 | 0.066 | 351.797 | 10.42 |
| id8i19grq | 2457938.70907 | 2457938.46477 | 26.68 | 26.36 | 0.15 | 43.32 | 42.31 | 0.065 | 351.797 | 10.36 |
| id8i19gtq | 2457938.71518 | 2457938.47088 | 26.93 | 26.61 | 0.15 | 43.32 | 42.31 | 0.065 | 351.797 | 10.60 |
| id8i20h2q | 2457938.75695 | 2457938.51265 | 26.86 | 26.54 | 0.14 | 43.32 | 42.31 | 0.065 | 351.797 | 10.53 |
| id8i20h3q | 2457938.76306 | 2457938.51876 | 26.70 | 26.38 | 0.14 | 43.32 | 42.31 | 0.065 | 351.797 | 10.37 |
| id8i20h5q | 2457938.76917 | 2457938.52487 | 26.53 | 26.21 | 0.14 | 43.32 | 42.31 | 0.065 | 351.797 | 10.20 |
| id8i20h7q | 2457938.77528 | 2457938.53098 | 26.84 | 26.52 | 0.14 | 43.32 | 42.31 | 0.064 | 351.797 | 10.51 |
| id8i20h9q | 2457938.78140 | 2457938.53710 | 26.89 | 26.57 | 0.14 | 43.32 | 42.31 | 0.064 | 351.797 | 10.56 |
| id8i21hbq | 2457938.82319 | 2457938.57889 | 27.27 | 26.95 | 0.17 | 43.32 | 42.31 | 0.064 | 351.797 | 10.94 |
| id8i21hcq | 2457938.82930 | 2457938.58500 | 26.81 | 26.49 | 0.17 | 43.32 | 42.31 | 0.064 | 351.797 | 10.49 |
| id8i21heq | 2457938.83541 | 2457938.59111 | 27.29 | 26.97 | 0.17 | 43.32 | 42.31 | 0.063 | 351.797 | 10.96 |
| id8i21hgq | 2457938.84152 | 2457938.59722 | 27.26 | 26.94 | 0.17 | 43.32 | 42.31 | 0.063 | 351.797 | 10.94 |
| id8i21hiq | 2457938.84763 | 2457938.60333 | 26.75 | 26.43 | 0.17 | 43.32 | 42.31 | 0.063 | 351.797 | 10.43 |
| id8i22hyq | 2457939.02184 | 2457938.77754 | 26.56 | 26.24 | 0.13 | 43.32 | 42.31 | 0.061 | 351.797 | 10.23 |
| id8i22hzq | 2457939.02795 | 2457938.78365 | 26.73 | 26.41 | 0.13 | 43.32 | 42.31 | 0.061 | 351.797 | 10.41 |
| id8i22i1q | 2457939.03406 | 2457938.78976 | 26.54 | 26.22 | 0.13 | 43.32 | 42.31 | 0.061 | 351.797 | 10.21 |
| id8i22i3q | 2457939.04017 | 2457938.79587 | 27.05 | 26.73 | 0.13 | 43.32 | 42.31 | 0.061 | 351.797 | 10.73 |
| id8i22i5q | 2457939.04628 | 2457938.80198 | 26.52 | 26.20 | 0.13 | 43.32 | 42.31 | 0.060 | 351.797 | 10.20 |
| id8i23i7q | 2457939.08806 | 2457938.84376 | 26.67 | 26.35 | 0.14 | 43.32 | 42.31 | 0.060 | 351.797 | 10.34 |
| id8i23i8q | 2457939.09417 | 2457938.84987 | 26.42 | 26.10 | 0.14 | 43.32 | 42.31 | 0.060 | 351.797 | 10.09 |
| id8i23iaq | 2457939.10028 | 2457938.85598 | 26.74 | 26.42 | 0.14 | 43.32 | 42.31 | 0.060 | 351.797 | 10.42 |
| id8i23icq | 2457939.10640 | 2457938.86210 | 26.97 | 26.65 | 0.14 | 43.32 | 42.31 | 0.060 | 351.797 | 10.64 |
| id8i23ieq | 2457939.11251 | 2457938.86821 | 27.10 | 26.78 | 0.14 | 43.32 | 42.31 | 0.060 | 351.797 | 10.78 |
| id8i24igq | 2457939.15428 | 2457938.90998 | 26.70 | 26.38 | 0.12 | 43.32 | 42.31 | 0.059 | 351.797 | 10.38 |
| id8i24ihq | 2457939.16039 | 2457938.91609 | 26.74 | 26.42 | 0.12 | 43.32 | 42.31 | 0.059 | 351.797 | 10.42 |
| id8i24ijq | 2457939.16650 | 2457938.92220 | 26.82 | 26.50 | 0.12 | 43.32 | 42.31 | 0.059 | 351.797 | 10.49 |
| id8i24ilq | 2457939.17261 | 2457938.92831 | 26.67 | 26.35 | 0.12 | 43.32 | 42.31 | 0.059 | 351.797 | 10.34 |
| id8i24inq | 2457939.17872 | 2457938.93442 | 26.98 | 26.66 | 0.12 | 43.32 | 42.31 | 0.059 | 351.797 | 10.65 |

## FIGURES

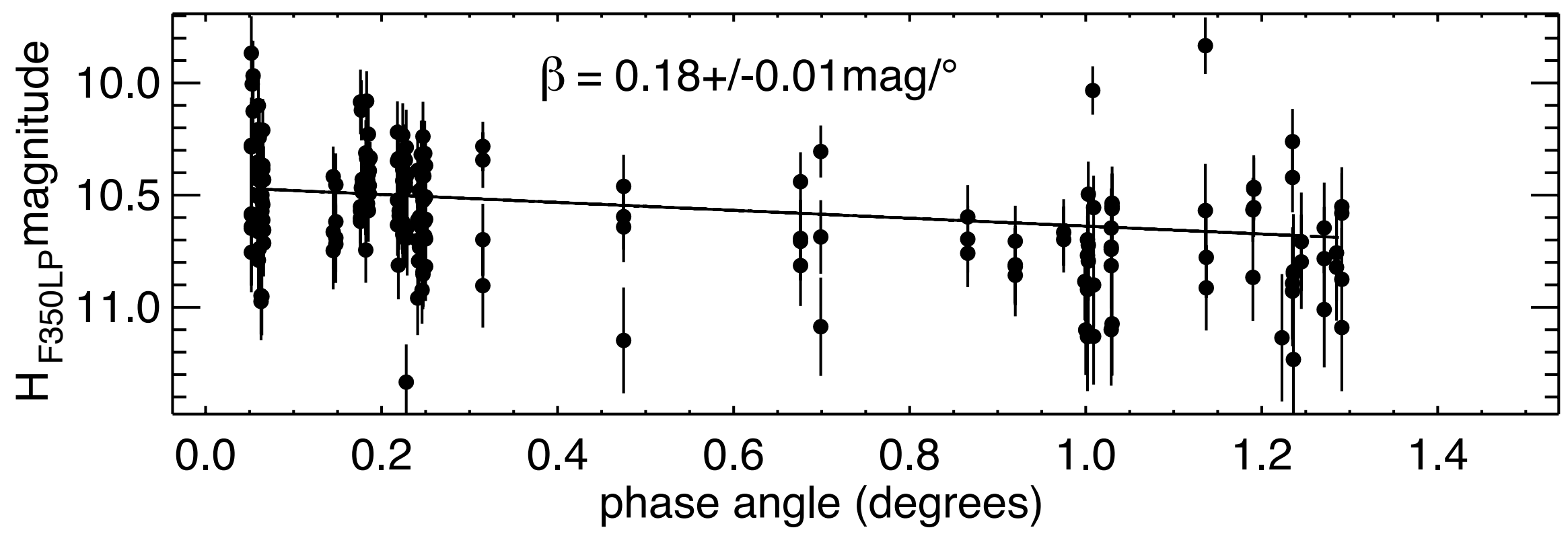

**Figure 1.** Linear fit to the non-phase, but geometrically corrected, data (1-sigma error bars are plotted) combining all of the measurements acquired over 4 years. The resulting phase coefficient is β = 0.18±0.01 mag/deg.

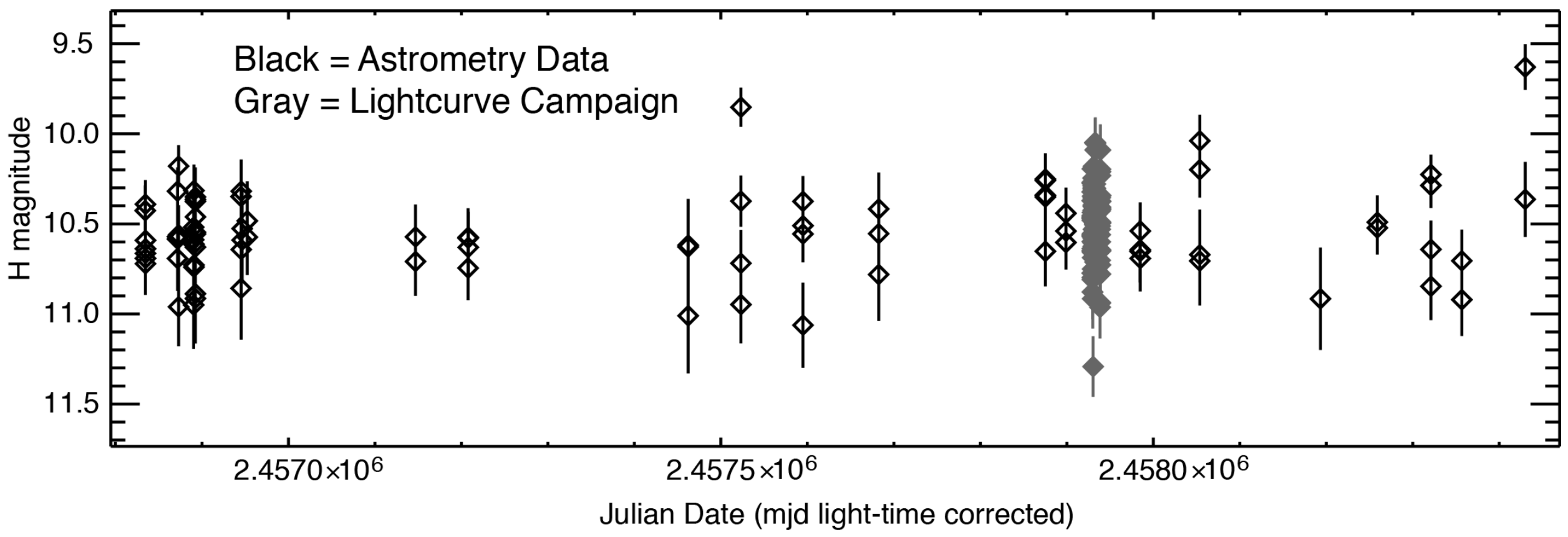

**Figure 2.** Geometrically corrected photometry from astrometric (black open points) and lightcurve (gray filled points) measurements 2014-2018 (1-sigma error bars plotted). The combined dataset covers 4 years while the lightcurve data are concentrated over 9.36 days. Although there is quite a bit of scatter in the data and we cannot determine the period, it would be possible to hide a lightcurve amplitude of up to 0.6 magnitudes within this dataset.

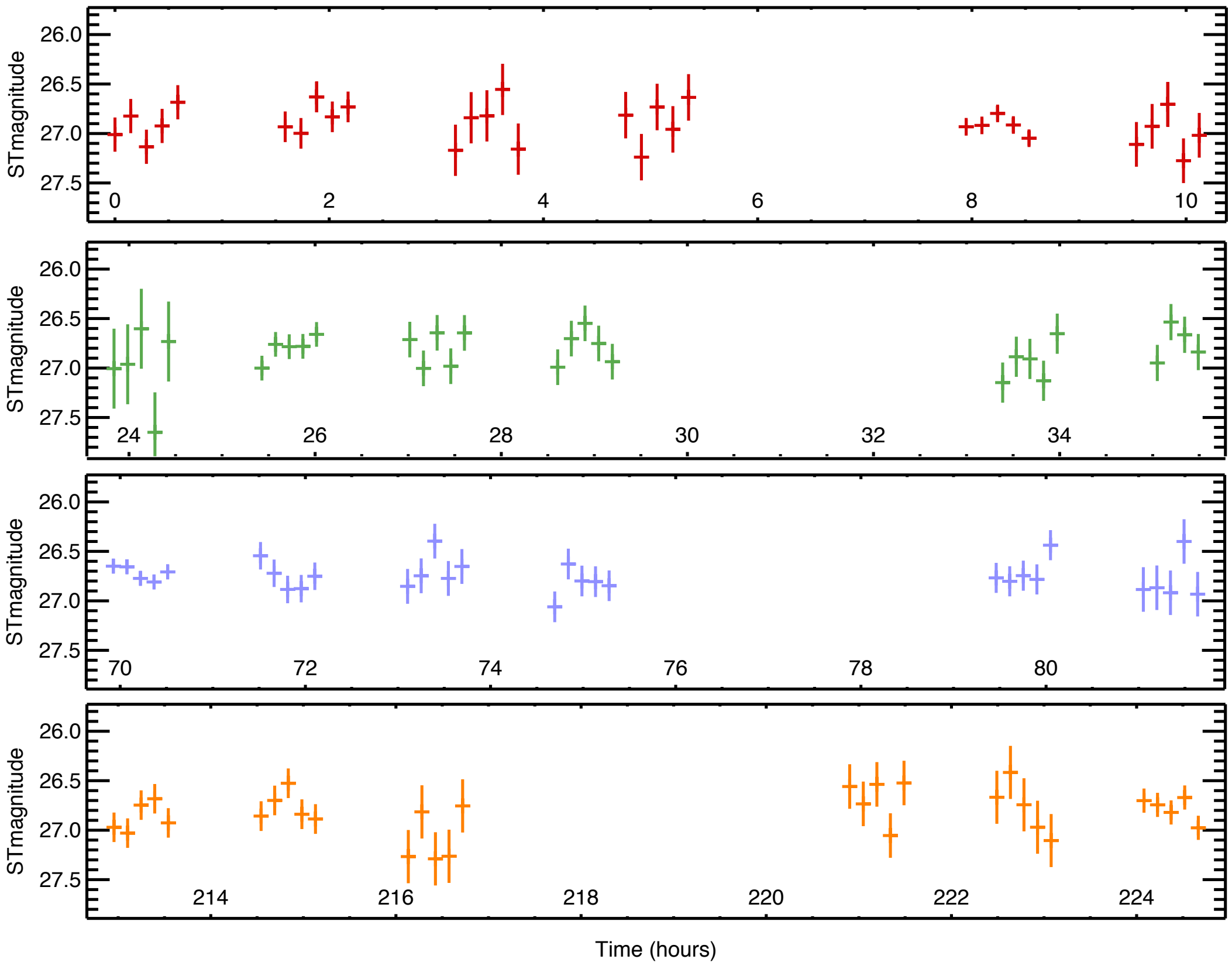

**Figure 3.** Individual data points for each lightcurve campaign visit (sets of 6 semi-consecutive HST orbits) with 1-sigma error bars. The approximate 2-orbit gap between the measurements in each set was due to a telescope constraint that required HST to be repointed after no more than 4 orbits. The gap does not severely impact the use of the data for our purposes.

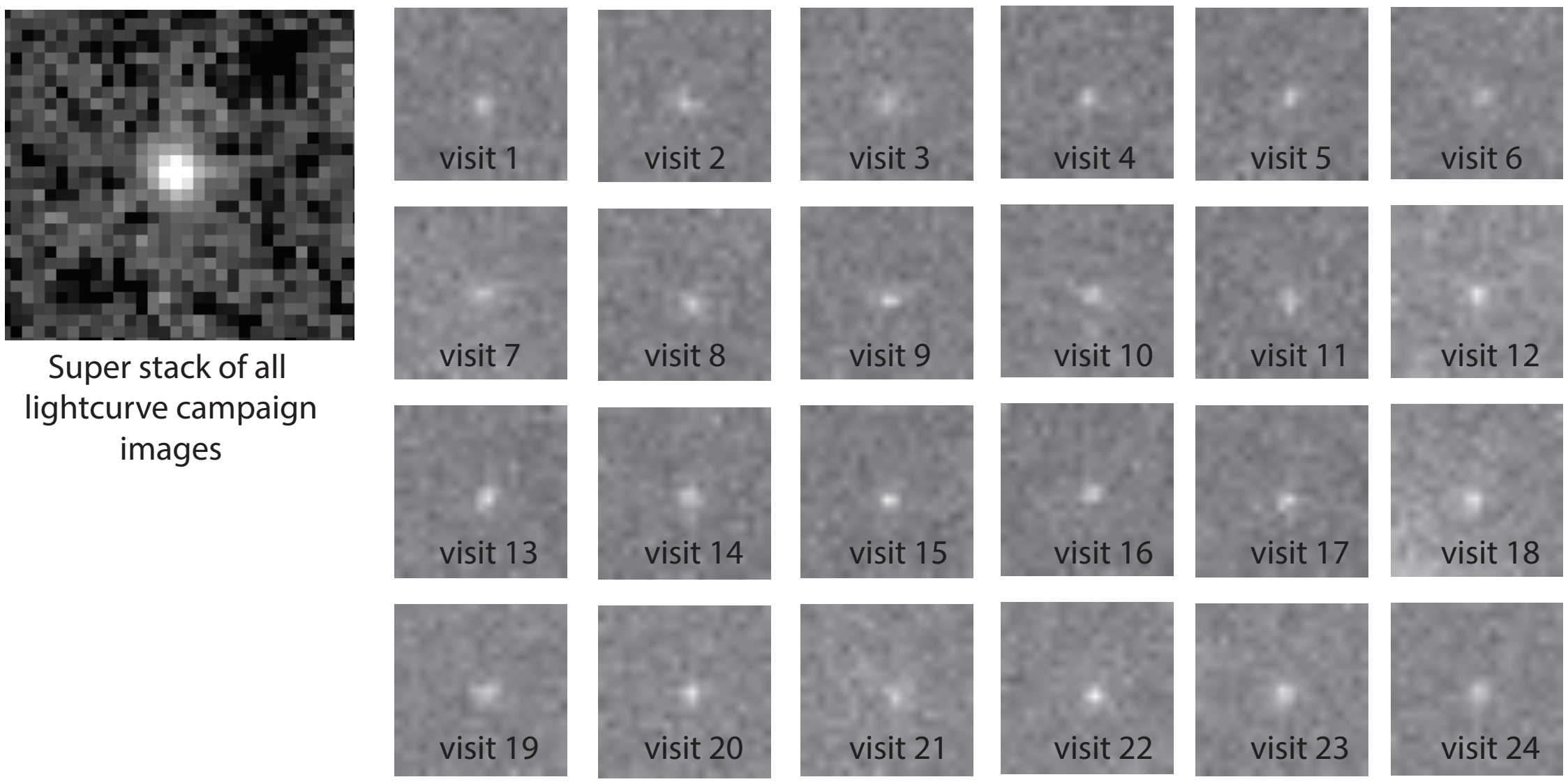

**Figure 4.** HST image stacks for each lightcurve visit and from the entire lightcurve campaign. The magnitude limit is ~29.0 in the F350LP filter. There are no obvious satellites in these images and attempts at binary PSF fits confirm the lack of a binary detection at the resolution of HST.

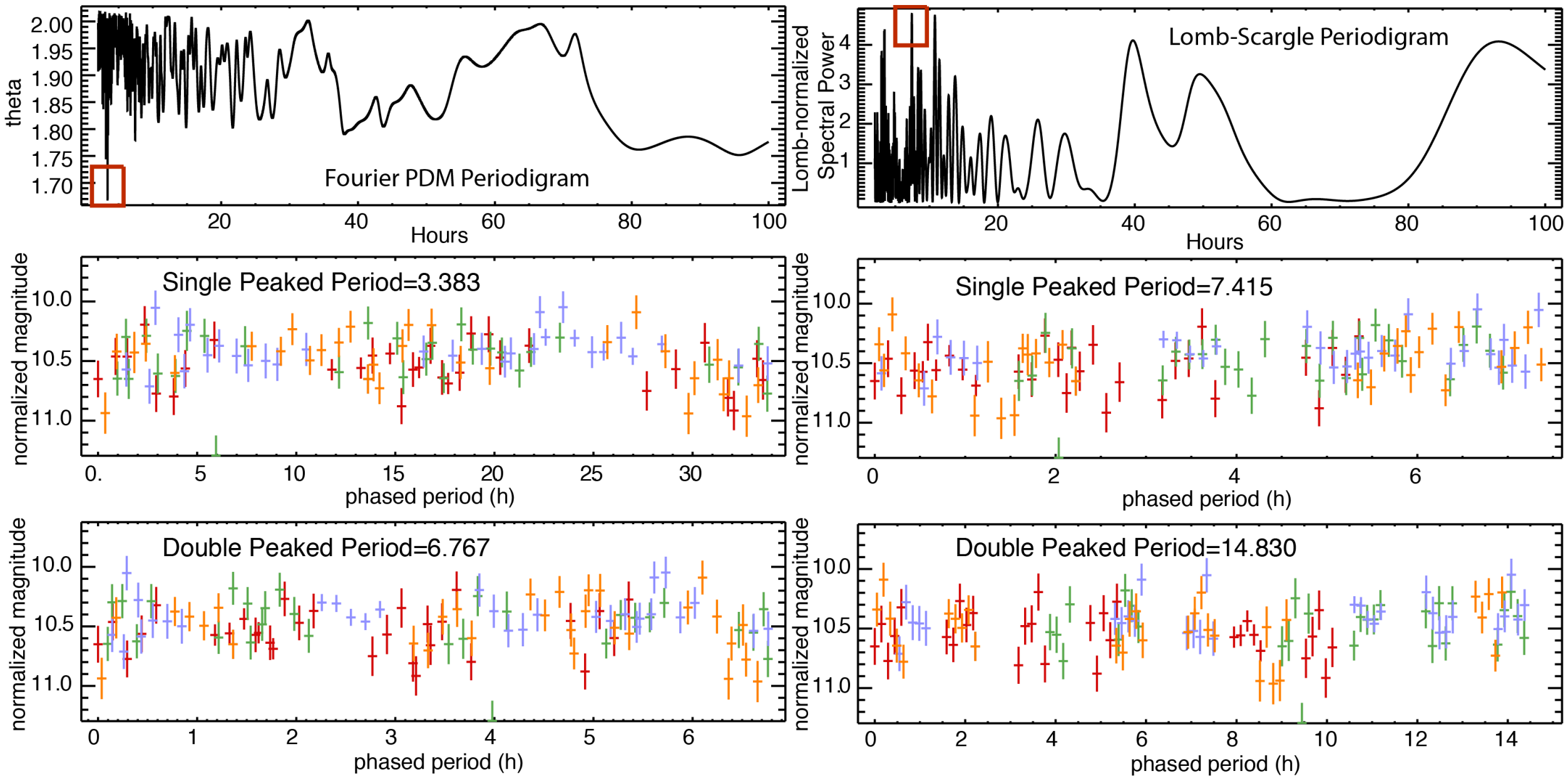

**Figure 5.** (left) Fourier PDM Periodigram and (right) Lomb-Scargle Periodigram. The most consistent period derived from the lightcurve campaign (looking from 2-10, 2-40 and 2-100 hours) using both search algorithms is a single-peaked period of 3.4 hours, double-peaked period of 6.8 hours. The Lomb-Scargle settles on a period of 7.4 (14.8 double-peaked) hours with a slightly stronger spectral power; however, in no case is one period found with statistical significance. The error bars (1-sigma in the plots) in both datasets are still quite large although the individual periods that come out of each individual analysis are somewhat consistent with each other.

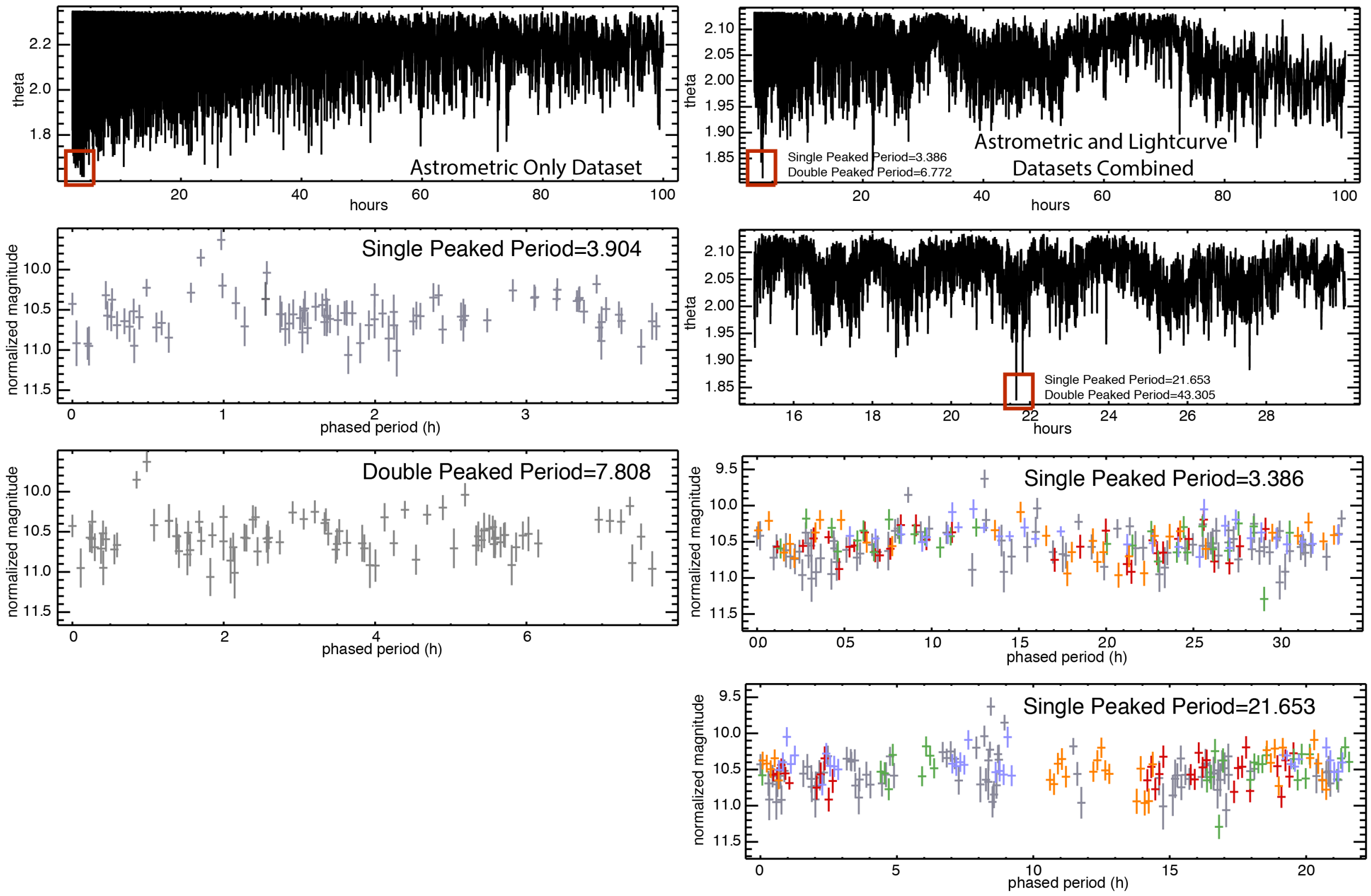

**Figure 6.** (left) Astrometric-only dataset and (right) combined astrometric and lightcurve datasets using the Fourier PDM algorithm (with 1-sigma error bars). In the astrometric-only dataset, periods near the short end dominate the periodogram results. In the combined datasets we nearly exclusively find a period of 3.38 hours, double-peaked period of 6.76 hours, although if we exclude periods less than 4 hours we find a period near 21.6 hours (43.2 hours). Since there are more data points in the lightcurve campaign then in the entire astrometric campaign this first dataset still dominates the conclusion, but we note that the astrometric campaign data do not appear to contradict the short period interpretation. In any case our results are not statistically significant.

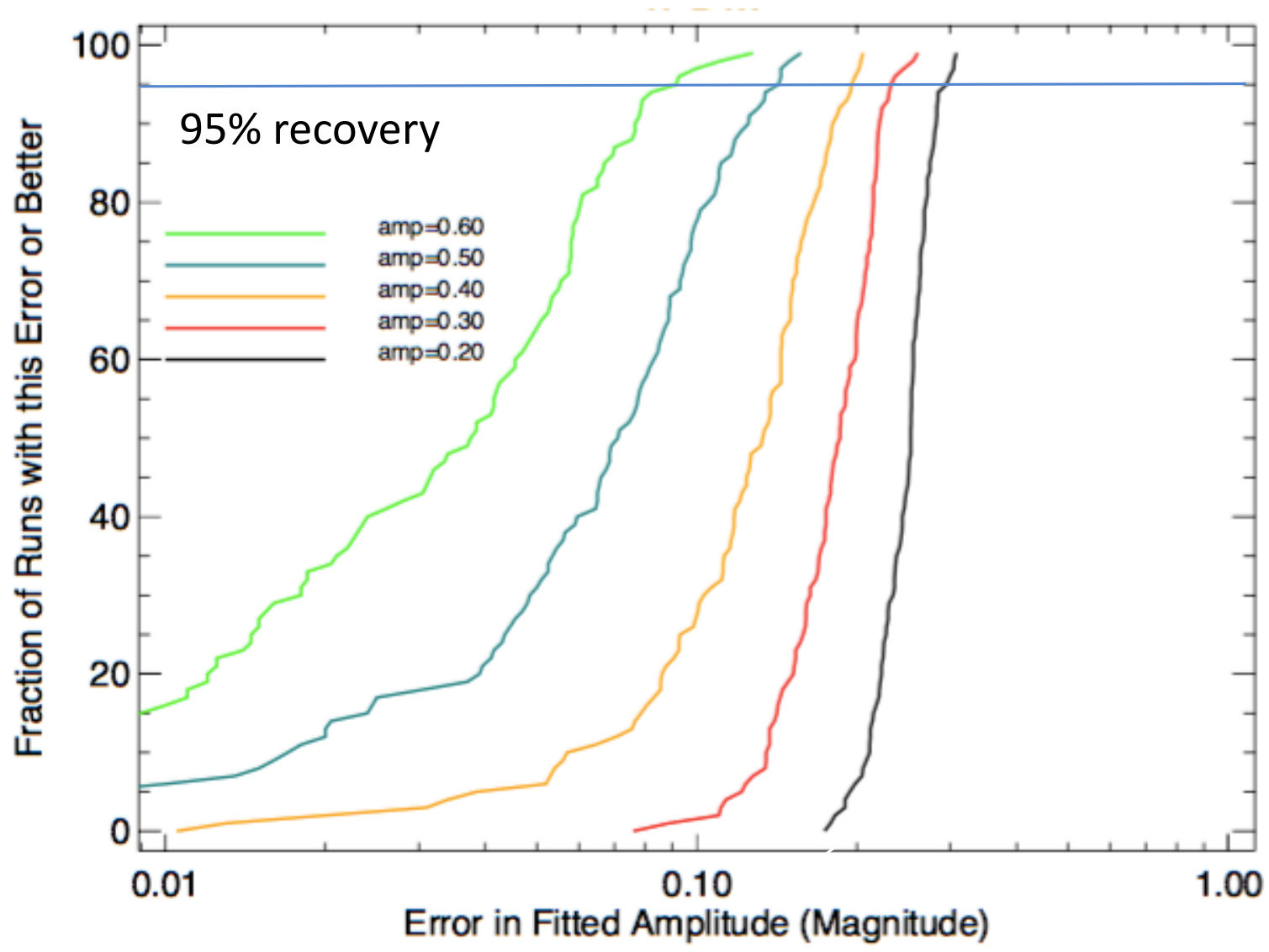

**Figure 7.** Synthetic analysis of potential lightcurve amplitudes in which we seed the HST dataset with rotation curves having amplitudes ranging from 0.2 to 0.6 magnitudes with similar uncertainties as we have in the HST lightcurve campaign dataset. The larger the amplitude the easier it is to make an accurate determination. Fundamentally this tells us that if MU69 was rotating rapidly with a large amplitude, then we should have been able to detect that given the S/N in our data.

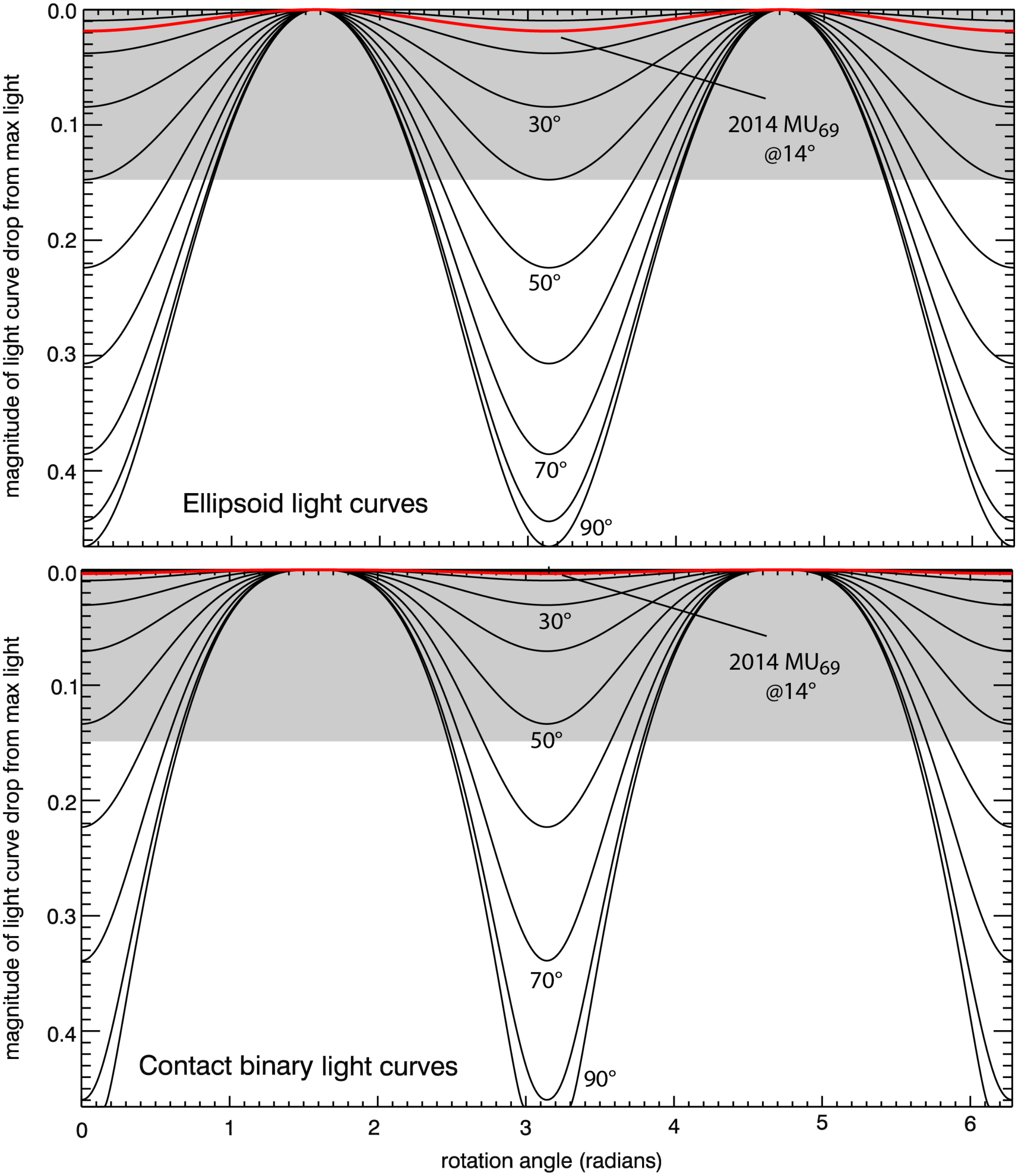

**Figure 8.** (Top): Model ellipsoidal lightcurve results for the full range of pole inclinations based on the work of Connolley & Ostro 1984. (Bottom) Model contact binary light curves for the full range of pole inclinations assuming both objects are completely spherical, there is no "neck" and that the light reflected from each object is exactly proportional to the visible object area. The 14° inclination of 2014 MU69's pole relative to the Earth is highlighted in red on each of these plots and changes by a factor of 6 between the two configurations. While these models are both simplistic, for a lightcurve amplitude as large as the scatter in our data, 0.15 magnitudes, to be produced would require a contact binary pole inclination of ≥52° or an ellipsoidal pole inclination of ≥40° (gray area in each plot).